\documentclass[%
 reprint,
 superscriptaddress,
nofootinbib,
 amsmath,amssymb,
 aps,
 prd,
 floatfix
]{revtex4-2}

\usepackage[T1]{fontenc}
\usepackage[utf8]{inputenc}
\usepackage{lmodern}
\usepackage{graphicx}
\usepackage{xcolor}
\usepackage{hyperref}

\usepackage{siunitx}

\newcommand{\F}{\mathcal{F}}
\newcommand{\A}{\mathcal{A}}
\newcommand{\M}{\mathcal{M}}
\newcommand{\Aplus}{A_{+}}
\newcommand{\Across}{A_{\times}}
\renewcommand{\S}{\mathcal{S}}
\newcommand{\dop}{\lambda}
\newcommand{\mle}{'}
\newcommand{\seg}{_{\mathrm{seg}}}
\newcommand{\Tspan}{T_{\mathrm{span}}}
\renewcommand{\det}{_{\mathrm{det}}}

\newcommand{\Tdata}{T_{\mathrm{data}}}
\newcommand{\av}[1]{\langle{#1}\rangle}
\newcommand{\avS}[1]{\langle\boldsymbol{#1}\rangle}
\newcommand{\dof}{\nu}  
\newcommand{\cosi}{\cos\iota}
\newcommand{\cosisq}{\cos^2\!\iota}
\renewcommand{\sc}[1]{\widehat{#1}}
\newcommand{\Fsc}{\sc{\F}}
\newcommand{\subA}{_{\!_{\mathrm{A}}}}
\newcommand{\subB}{_{\!_{\mathrm{B}}}}
\newcommand{\subAB}{_{\!_{\mathrm{AB}}}}

\newcommand{\subc}{_{\mathrm{c}}}
\newcommand{\subs}{_{\mathrm{s}}}
\newcommand{\subsc}{_{\mathrm{s,c}}}
\newcommand{\subphi}{_{\phi}}

\newcommand{\fa}{_{\mathrm{fa}}}

\newcommand{\prob}[2]{P\left(#1|#2\right)}

\newcommand{\stat}{\mathfrak{s}}
\newcommand{\erfc}{\mathrm{erfc}}
\newcommand{\cond}[1]{\mathrm{cond}[#1]}
\newcommand{\depth}{\mathcal{D}}
\newcommand{\udepth}{\si{1\per\sqrt\hertz}}

\usepackage{soul}

\newcommand{\commitDATE}{2022-04-25 17:06:27 +0200}

\newcommand{\commitIDshort}{commitID: bbdc975}
\newcommand{\commitSTATUS}{CLEAN}

\begin{document}
\preprint{APS/123-QED}

\title{Improved short-segment detection statistic for continuous gravitational waves}
\author{%
P.~B.~Covas$^{1}$, R. Prix$^{1}$
}\noaffiliation
\affiliation{Max Planck Institute for Gravitational Physics (Albert Einstein Institute), D-30167 Hannover, Germany}
\date{\commitDATE; \commitIDshort-\commitSTATUS}

\begin{abstract}
  Continuous gravitational waves represent one of the long-sought types of
  signals that have yet to be detected. Due to their small amplitude, long
  observational datasets (months-years) have to be analyzed together, thereby
  vastly increasing the computational cost of these searches. All-sky searches
  face the most severe computational obstacles, especially searches for sources
  in unknown binary systems, which need to break the data into very short
  segments in order to be computationally feasible. In this paper, we present a
  new detection statistic that improves sensitivity by up to \SI{19}{\percent}
  compared to the standard $\F$-statistic for segments shorter than a few hours.
\end{abstract}

\maketitle

\section{Introduction}
\label{sec:introduction}

Continuous gravitational waves (CWs) are long-lasting and nearly monochromatic
signals, expected to be emitted by asymmetric rotating neutron stars
\cite{prix06:_cw_review, ReviewCWIso}.
While there have been many searches for CWs performed to date, none has yet been able to make a detection.
One of the main difficulties in finding CWs is their small amplitude, expected
to be many orders of magnitude below the noise floors of current detectors.
In order to accumulate a detectable signal-to-noise ratio, large datasets
(spanning months--years) therefore have to be combined in a search, which
complicates the analysis and results in a vast increase in computing cost due to
the astronomically large number of templates to be searched.

We can divide searches for CWs into three main categories, ordered by increasing
computational cost: (i) targeted searches, (ii) directed searches, and (iii)
all-sky searches.
Directed and all-sky searches typically cannot combine all the data coherently
due to the unfeasible computational cost this would entail, and instead have to
employ semi-coherent algorithms by breaking the data into shorter segments
(e.g., see \cite{brady_searching_2000,PrixShaltev2011..optimalStackSlide}).
Although the total signal power does not depend on the number of segments used,
the background (i.e., noise) distribution worsens when the number of segments
increases.
Therefore, while using more segments dramatically alleviates the computational
cost of the search, it also reduces the resulting sensitivity.
The optimal search setup consists of using the longest segments possible within
a given computational cost budget \cite{PrixShaltev2011..optimalStackSlide}.

There are various ways to compute semi-coherent detection statistics, but one
widely-used approach we focus on here consists in summing coherent
$\F$-statistics \cite{jks98:_data,cutler05:_gen_fstat} across segments, known as
the \emph{StackSlide} approach (see also \cite{2018arXiv180802459D} for more
discussion and an overview of different methods currently in use).
The corresponding $\F$-statistic segments used have so far always been longer than
$\sim\SI{11}{\hour}$.
At the other end of the spectrum are the ``power'' statistics, which directly
use Fourier power over short segments, typically no longer than
$\sim{\SI{1}{\hour}}$, as the per-segment statistic.

A recent search \cite{covas_constraints_2022} has bridged this gap for the first
time by employing a semi-coherent $\F$-statistic on short segments of
$\SI{900}{\second}$, designed to improve the sensitivity of an all-sky search
for signals from unknown neutron stars in binary systems.
Using the $\F$-statistic on such short segments has resulted in unexpected and
previously-unknown numerical difficulties, with the $\F$-statistic becoming
singular in many segments, especially those containing data from only one
detector.
This degeneracy can be easily understood from the underlying antenna-pattern
matrix becoming ill-conditioned for short coherence times and the $\F$-statistic
relying on inverting this matrix.
Intuitively this corresponds to a failure of the maximum-likelihood estimation
of the four unknown amplitude parameters (describing the ``$+$'' and
``$\times$'' polarizations) for short coherence times, during which the
detector hardly moves.

In an attempt to fix this singularity by constructing a well-behaved
``fallback'' statistic we have discovered a new detection statistic that is not
just well-behaved in the short-segment limit but turns out to be \emph{more
  sensitive} than the $\F$-statistic for segments shorter than a few hours, even
when the $\F$-statistic is far from numerically singular.

This paper is organized as follows: in Sec.~\ref{sec:standard-f-statistic} we
introduce the standard $\F$-statistic; Sec.~\ref{sec:degen-short-segm} discusses
the singular limit of the $\F$-statistic for short segments, and
Sec.~\ref{sec:impr-short-segm} introduces the new statistic. Section
\ref{sec:impr-short-segm} characterizes the sensitivity improvement of the new
statistic compared to the $\F$-statistic, and Sec.~\ref{sec:numerical-tests}
provides numerical tests for these analytical results and further characterizes
the new statistic, followed by conclusions in Sec.~\ref{sec:conclusions}.
Appendix \ref{subsubsec:constantantenna} discusses an alternative short-segment
statistic construction that turns out to be unsuitable for multi-detector
setups.

\section{The Standard $\F$-statistic}
\label{sec:standard-f-statistic}

\subsection{The continuous-waves likelihood}
\label{sec:cw-likelihood}

We can parameterize the gravitational-wave signal from a non-axisymmetric
rotating neutron star by four amplitude parameters $\A$ and several
phase-evolution parameters $\dop$.
The four amplitude parameters consist of the overall signal amplitude $h_0$, the
inclination angle $\iota$ between the line of sight and the neutron star
rotation axis, the phase $\phi_0$ at a reference time, and a polarization angle $\psi$.
The phase-evolution parameters consist of the frequency of the signal $f$
(slowly changing over time), the sky position of the neutron star, and
binary-orbital parameters if the neutron star is in a binary.

The CW signal depends non-linearly on the physical amplitude parameters
$\{h_0,\cosi,\psi,\phi_0\}$, but in \cite{jks98:_data} the authors found a set
of four amplitude coordinates $\A^\mu$ that linearize the functional form of the
signal, namely
\begin{equation}
  \begin{aligned}
    \A^{1} &= \Aplus \cos \phi_{0} \cos 2 \psi - \Across \sin \phi_{0} \sin 2 \psi\,, \\
    \A^{2} &= \Aplus \cos \phi_{0} \sin 2 \psi + \Across \sin \phi_{0} \cos 2 \psi\,, \\
    \A^{3} &= -\Aplus \sin \phi_{0} \cos 2 \psi - \Across \cos \phi_{0} \sin 2 \psi\,, \\
    \A^{4} &= -\Aplus \sin \phi_{0} \sin 2 \psi + \Across \cos \phi_{0} \cos 2 \psi\,,
  \end{aligned}
\end{equation}
where $\Aplus = 0.5\, h_0\, ( 1 + \cos^2{\iota})$ and $\Across = h_0 \cosi$ are
the amplitudes of the ``$+$'' and ``$\times$'' polarization, respectively.
This allows one to write a signal $s^X(t)$ in the frame of detector $X$ as
\begin{equation}
  \label{eq:49}
  s^X(t;\A,\dop) = \sum_{\mu=1}^4 \A^{\mu}\, h^X_{\mu} (t; \dop),
\end{equation}
where the matched-filter basis functions are defined as
\begin{equation}
  \label{eq:basisfunc}
  \begin{aligned}
    h_{1}^{X}&\equiv a^{X}(t) \, \cos \phi^X(t),\quad h_{2}^{X} \equiv b^{X}(t) \,\cos \phi^X(t), \\
    h_{3}^{X}&\equiv a^{X}(t) \, \sin \phi^X(t), \quad  h_{4}^{X} \equiv b^{X}(t)\,\sin \phi^X(t),
  \end{aligned}
\end{equation}
in terms of the detector-frame signal phase $\phi^X(t)$ at time $t$, and
the antenna-pattern functions $a^X(t)$ and $b^X(t)$ \cite{jks98:_data,prix:_cfsv2}.

A detection statistic typically aims to distinguish two basic hypotheses about the
data $x(t)$:
(i) the data only consists of Gaussian noise $n$, i.e., $x=n$, or
(ii) the data consists of an astrophysical signal $s$ in addition to Gaussian noise, i.e.,
$x = n + s$.
The likelihood ratio $\mathcal{L}$ between these two hypothesis is
\begin{equation}
  \label{eq:loglikelihood}
  \begin{aligned}
    \ln \mathcal{L} (x; \A, \dop) &\equiv \ln \frac{\prob{x}{\text{signal}}}{\prob{x}{\text{noise}}}
    = (x | h)-\frac{1}{2}(h | h)\\
    & =\A^{\mu} x_{\mu}-\frac{1}{2} \A^{\mu} \M_{\mu \nu} \A^{\nu}\,,
  \end{aligned}
\end{equation}
with implicit summation over $\mu,\nu=1,\ldots,4$ and where we defined
\begin{equation}
  \label{eq:xmu}
  x_{\mu} \equiv (x|h_\mu),\quad
  \M_{\mu\nu} \equiv (h_\mu|h_\nu)\,,
\end{equation}
in terms of the multi-detector scalar product\footnote{For simplicity of
  notation here we assume stationary noise and contiguous data $T$, see
  \cite{prix:_cfsv2} for a more general expression.}\cite{cutler05:_gen_fstat,prix:_cfsv2}
\begin{equation}
  \label{eq:4}
  (x|y) \equiv 2 \S^{-1} \sum_{X}^{N\det} \sqrt{w_{X}} \int_{0}^{T} x^{X}(t)\, y^X(t)\, dt,
\end{equation}
where $\S$ represents the overall noise floor, defined as
\begin{equation}
  \label{eq:54}
  \S^{-1} \equiv \frac{1}{N\det} \sum_{X} S_{X}^{-1},
\end{equation}
and $w_X$ is a per-detector noise weight defined as
\begin{equation}
  \label{eq:5}
  w_X \equiv \frac{S_X^{-1}}{\S^{-1}}\,.
\end{equation}
For ease of notation, here we assume stationary noise floors $S_X$ for each
detector $X$, while the numerical implementation
\cite{prix:_cfsv2,prix06:_searc} uses a more general formulation valid also for
slowly-varying noise-floors.

The antenna-pattern matrix $\M_{\mu\nu}$ defined in Eq.~\eqref{eq:xmu} can be
expressed more explicitly as
\begin{equation}
  \label{eq:14}
  \M_{\mu\nu} = \gamma \begin{pmatrix}
    A & C & 0 & 0 \\
    C & B & 0 & 0 \\
    0 & 0 & A & C \\
    0 & 0 & C & B
  \end{pmatrix},
\end{equation}
with the ``data factor'' $\gamma$ defined as
\begin{equation}
  \label{eq:2}
  \gamma \equiv \S^{-1}\,\Tdata,
\end{equation}
where $\Tdata$ is the total amount of data over all detectors (e.g.,
$\Tdata=N\det\, T$ if there are no data gaps) and the sky-position dependent
antenna-pattern coefficients
\begin{equation}
  \label{eq:APC}
  A \equiv \avS{a^2}\,,\quad
  B \equiv \avS{b^2},\quad
  C \equiv \avS{a\,b}\,,
\end{equation}
using the noise-weighted multi-detector time average
\begin{equation}
  \label{eq:1}
  \begin{aligned}
    \avS{Q} &\equiv \frac{1}{N\det} \sum_{X}^{N\det} \sqrt{w_{X}}\, \av{Q^X}\,,\quad\text{with}\\
    \av{Q^X} &\equiv \frac{1}{T}\int_0^T  Q^{X}(t)\,dt\,.
  \end{aligned}
\end{equation}
We define the block determinant $D$ of $\M$ as
\begin{equation}
  \label{eq:16}
  D \equiv A\,B - C^2\,,
\end{equation}
such that $\mathrm{det}\,\M = D^2$.
Note that the $x_\mu=n_\mu+s_\mu$ in Eq.~\eqref{eq:xmu} are
Gaussian distributed with expectation and second moment
\begin{equation}
  \label{eq:20}
  E\left[x_\mu\right] = s_\mu\,,\quad
  E\left[x_\mu\,x_\nu\right] = \M_{\mu\nu} + s_\mu\,s_\nu\,.
\end{equation}

\subsection{The coherent $\F$-statistic}
\label{sec:coherent-f-statistic}

The likelihood ratio in Eq.~\eqref{eq:loglikelihood} can be analytically maximized
over the four amplitude coordinates $\A^\mu$, yielding the maximum-likelihood estimates
\begin{equation}
  \label{eq:13}
  \A\mle^{\mu} = \M^{\mu\nu}\, x_{\nu}\,,
\end{equation}
where $\M^{\mu\nu}$ is the inverse of the antenna-pattern matrix. Substituting
back into Eq.~\eqref{eq:loglikelihood} defines the $\F$-statistic as
\begin{equation}
    2 \F(x;\dop) \equiv \ln\mathcal{L}(x;\dop;\A\mle) = x_{\mu}\,\M^{\mu\nu} x_{\nu}\,.
    \label{eq:2F}
  \end{equation}
It will be useful to write this out more explicitly in terms of the
antenna-pattern-matrix elements of Eq.~\eqref{eq:14} as
\begin{equation}
  \label{eq:15}
  2\F = \frac{(x_1^2 + x_3^2)B + (x_2^2 + x_4^2)A - 2(x_1 x_2 + x_3 x_4)C}{\gamma\,D}.
\end{equation}
An alternative derivation of the $\F$-statistic as a Bayes factor
\cite{2009CQGra..26t4013P} reveals the underlying amplitude priors to be
unphysical, which is why the $\F$-statistic is not statistically optimal.
The main advantage of the $\F$-statistic is the analytical elimination of the
four amplitude parameters, which gives it a computational advantage over any
alternative that would require explicit numerical operations to deal with the
unknown amplitude parameters (e.g., see
\cite{2014CQGra..31f5002W,wetteAnalitycal,2019CQGra..36a5013B} for further
discussion).

Another useful property of the $\F$-statistic is its known $\chi^2$ probability
distribution with $\dof=4$ degrees of freedom and noncentrality parameter
\begin{equation}
  \label{eq:19}
  \rho^2\equiv (s|s) = \A^\mu\,\M_{\mu\nu}\A^\nu\,.
\end{equation}
The known mean and variance of this distribution are
\begin{equation}
  \label{eq:28}
  \begin{aligned}
    E[2\F] &= \dof + \rho^2,\\
    \text{var}[2\F] &= 2(\dof + 2\rho^2)\,.
  \end{aligned}
\end{equation}
It will be useful to express the noncentrality as
\begin{equation}
  \label{eq:rho2}
  \rho^{2} = h_{0}^{2} \,\gamma\,\left(\alpha_{1}\,A + \alpha_{2}\,B + 2 \alpha_{3}\,C \right),
\end{equation}
in terms of amplitude angle factors $\alpha_i(\cosi,\psi)$ defined as
\begin{equation}
  \label{eq:51}
  \begin{aligned}
    \alpha_{1} &\equiv \frac{1}{4}\left(1+\cosisq\right)^{2} \cos ^{2}2\psi+\cosisq\,\sin ^{2}2 \psi \\
    \alpha_{2} &\equiv \frac{1}{4}\left(1+\cosisq\right)^{2} \sin ^{2}2\psi+\cosisq\,\cos ^{2}2 \psi \\
    \alpha_{3} &\equiv \frac{1}{4}\left(1-\cosisq\right)^{2} \sin 2 \psi\, \cos 2 \psi.
  \end{aligned}
\end{equation}
One can see that averaging over the unknown $\cosi$ and $\psi$ yields
$\av{\alpha_1}_{\cosi,\psi} = \av{\alpha_2}_{\cosi,\psi}=2/5$
and $\av{\alpha_3}_{\cosi,\psi}=0$, and therefore the angle-averaged
noncentrality parameter can be found as
\begin{equation}
  \av{\rho^2}_{\cosi,\psi} = \frac{2}{5}\,h_0^2\,\gamma\, ( A + B ).
  \label{eq:rho2Av}
\end{equation}

\subsection{The semi-coherent $\Fsc$-statistic}
\label{sec:semi-coherent-f}

The use of the coherent $\F$-statistic as a search method is practically limited
to very small parameter spaces, due to the rapidly-growing computing cost of a
coherent template bank with increasing observation times $T$.
Wide-parameter-space searches therefore have to employ cheaper
\emph{semi-coherent} methods, which typically result in better sensitivity at
constrained computing cost \cite{brady_searching_2000}.
In the following we focus on the semi-coherent version of the $\F$-statistic.

We can break the total observation time $T$ into $\ell = 1,\ldots, N\seg$
\emph{segments} of duration $T\seg$, and define the semi-coherent
$\Fsc$-statistic as
\begin{equation}
  \label{eq:10}
  \Fsc \equiv \sum_{\ell=1}^{N\seg} \F_\ell\,,
\end{equation}
where $\F_\ell$ is the coherent $\F$-statistic computed on segment $\ell$.
Because this is a sum of $\chi^2$-distributed statistics, the semi-coherent
$\Fsc$-statistic is also $\chi^2$-distributed, with $\sc{\dof}$ degrees
of freedom and noncentrality parameter $\sc{\rho}^2$, given by
\begin{equation}
  \label{eq:12}
  \begin{aligned}
    \sc{\dof} &=\sum_\ell \dof_\ell = 4\,N\seg\,,\\
    \sc{\rho}^2 &=\sum_\ell\rho^2_\ell\,.
  \end{aligned}
\end{equation}
The optimal choice for the number and length of segments is subject to a
computing-cost-constrained optimization problem
\cite{PrixShaltev2011..optimalStackSlide,2016PhRvD..93f4011M,2018PhRvD..97b4051M},
but the general trend is that the larger the parameter space, the
shorter the segments must be in order for the search to be computationally
feasible.

For example, a recent all-sky search for continuous waves from neutron stars
in unknown binary systems \cite{covas_constraints_2022} had to use very short
segments of $T\seg=\SI{900}{\second}$ over an observation time of $T=\SI{6}{months}$.
This has revealed previously-unknown problems and limitations of the
$\F$-statistic over such short baselines, discussed in the next section.

\section{Problems with short-segment $\F$}
\label{sec:degen-short-segm}

\subsection{The ill-conditioned short-segment limit}

The $\F$-statistic relies on the inverse antenna-pattern matrix $\M^{\mu\nu}$,
as seen in Eq.~\eqref{eq:2F}.
However, for a single detector the short-duration limit of the matrix
coefficients $A$, $B$ and $C$ of Eq.~\eqref{eq:APC} is
\begin{equation}
  \label{eq:3}
  \{A, B, C\}\overset{T\seg\rightarrow0}{\longrightarrow} \{a^2, b^2, ab\},
\end{equation}
which leads to a singular antenna-pattern matrix with determinant
$\mathrm{det}\,\M = D^2 = \left(A\,B - C^2\right)^2\rightarrow 0$.
For short segments the matrix therefore becomes ill-conditioned and
computing $D$ is unreliable or fails.
For performance reasons the \textit{lalsuite}\cite{lalsuite} codes computing the
$\F$-statistic use single-precision for the antenna-pattern matrix, and
computing $D$ can therefore become problematic already at condition numbers above
$\cond{\M}\gtrsim\num{e4}$, for which the code will refuse to compute a result.

The problem for single-detector segments is illustrated in Fig.~\ref{fig:PercentageBadSegments}:
\begin{figure*}
  \centering
  \includegraphics[width=\columnwidth]{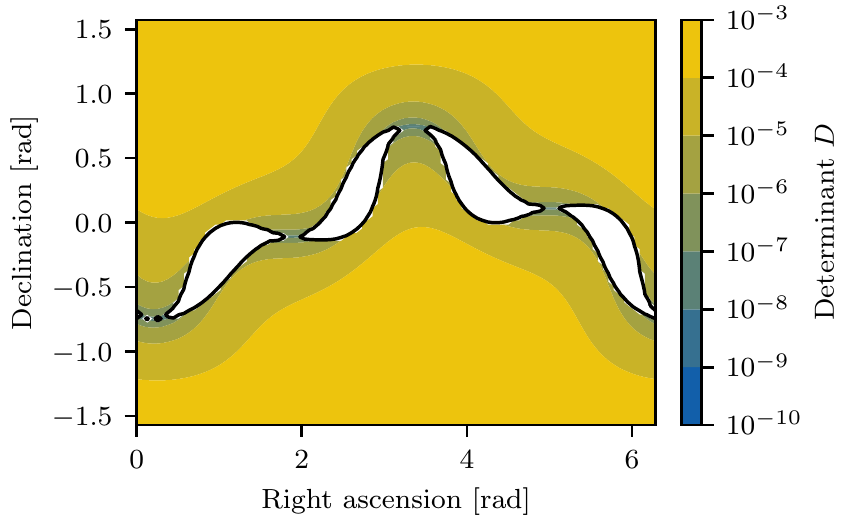}
  \includegraphics[width=\columnwidth]{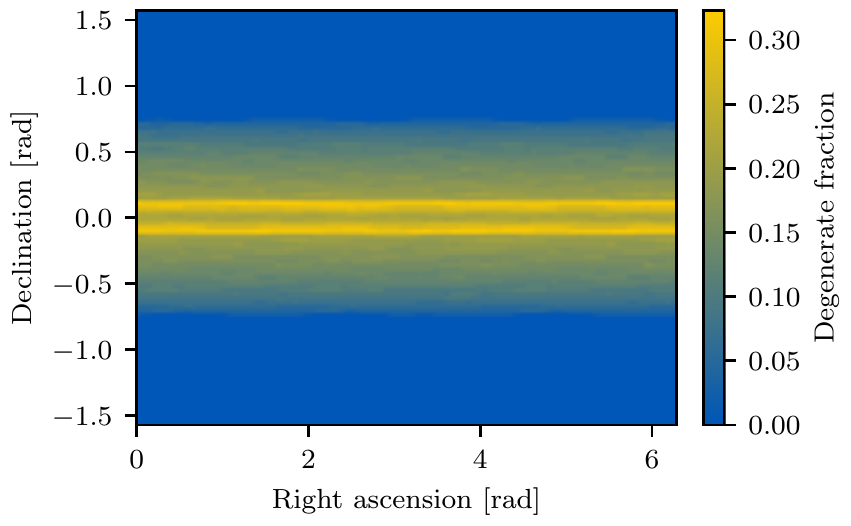}
  \caption{Antenna-pattern degeneracy for a single detector (H1) and short
    segments of $T\seg=\SI{900}{\second}$.
    \emph{Left plot:} block determinant $D$ over the sky for a single segment.
    Contour lines mark the region where $\M$ is considered numerically
    ill-conditioned, i.e.\ $\cond{\M}>\num{e4}$.
    \emph{Right plot:} fraction of segments with ill-conditioned antenna-pattern
    matrix as a function of sky position, using $N\seg=96$ segments.
    }
  \label{fig:PercentageBadSegments}
\end{figure*}
the left plot shows the block determinant $D$ and the critical condition-number
contour $\cond{\M}=\num{e4}$ over the sky for a single segment of $T\seg=\SI{900}{\second}$.
The right plot shows the fraction of segments with super-critical condition
number over the sky.
We see that near the equator $\sim\SI{40}{\percent}$ of the segments would
be ill-conditioned.

Combining data from more than one detector generally alleviates this problem, as
$C$ now tends to $\propto(\sum_X a^X)(\sum_Y b^Y)$, where the cross-terms
generally prevent the determinant from vanishing even when $T\rightarrow0$.
In practice, however, due to commonly-found gaps in the detector data, a
realistic segment setup with $T\seg=\SI{900}{\second}$ segments for two
detectors (H1+L1) (such as \cite{covas_constraints_2022}) will contain many
segments containing data from only one detector.
Therefore computing the $\F$-statistic on such segments will still encounter
numerical problems.

Dropping such segments would reduce the amount of usable data and therefore
sacrifice sensitivity (e.g., for the O3a LIGO science run, this affects about a
\emph{third} of such short segments \cite{covas_constraints_2022}).
This problem has motivated an investigation into a non-singular ``fallback''
option for the singular $\F$-statistic on short segments.

\subsection{Fallback construction for singular $\F$-statistic}
\label{sec:short-segm-fallb}

As discussed in the previous sections, the $\F$-statistic becomes singular in
the limit $T\seg\rightarrow 0$ for single-detector segments due to due failure
of maximum-likelihood estimation for all four amplitude parameters $\A^\mu$ that
describe the full ``$+$'' and ``$\times$'' polarization content of the CW.
These four amplitudes translate into $\dof=4$ degrees of freedom of the
resulting $\chi^2$ distribution governing the $\F$-statistic.

These considerations suggest a strategy for avoiding degeneracy by reducing the
number of amplitude parameters (i.e., degrees of freedom) we try to infer about
the signal: even with a single stationary detector, \emph{two} amplitudes should
always be perfectly well-determined, namely those corresponding to the ``$+$''
polarization aligned with the arms of the detector.
In the $\F$-statistic framework, the two antenna-pattern functions $a(t)$ and
$b(t)$ represent the responses to the two GW polarizations projected on a
sky-position dependent coordinate frame \cite{2014CQGra..31f5002W}.

Guided by this intuition, we construct a new detection statistic by selecting
the ``dominant response'' between $a$ and $b$ in each sky position.
In other words, we implicitly use only two well-determined amplitude parameters
and neglect the other two, effectively reducing the $\chi^2$ degrees of freedom
of the resulting statistic to $\dof=2$\footnote{A somewhat-related $\dof=2$
  degrees-of-freedom variant is the $\mathcal{G}$-statistic\cite{FstatGstat},
  which assumes that the two amplitude parameters $\cosi$ and $\psi$ are
  \emph{known} from observations for a given pulsar.}.

An interesting alternative construction (described in
Appendix~\ref{subsubsec:constantantenna}) consists in completely neglecting the
antenna-pattern response, resulting in a pure ``demodulated power'' statistic.
While this works for a single detector, it does not generalize well to multiple
detectors (suffering destructive interference), contrary to the
dominant-response statistic introduced here.

\subsubsection{The dominant-response $\F\subAB$-statistic}
\label{sec:domin-polar-f}

Let us consider the special case $b(t)= 0$, which implies $B=C=0$, $h_2=h_4=0$
and further $x_2=x_4=0$, such that the likelihood Eq.~\eqref{eq:loglikelihood}
reduces to
\begin{equation}
  \label{eq:18}
  \ln\mathcal{L}\subA = \A^1\,x_1 + \A^3\,x_3 -
  \frac{1}{2}\gamma\,A\left[(\A^1)^2 + (\A^3)^2\right]\,.
\end{equation}
Maximization over the remaining two amplitude coordinates $\A^1$ and $\A^3$ yields
\begin{equation}
  \label{eq:11}
  2\F\subA = \frac{x_1^2 + x_3^2}{\gamma A}\,.
\end{equation}
Similarly, if had assumed $a(t)=0$ instead, which implies $x_1=x_3=0$, we would find
\begin{equation}
  \label{eq:17}
  2\F\subB = \frac{x_2^2 + x_4^2}{\gamma\,B}\,.
\end{equation}
Given the noise expectations of Eq.~\eqref{eq:20}, in particular
$E\left[n_1^2\right] = E\left[n_3^2\right] = \gamma\,A$ and
$E\left[n_2^2\right] = E\left[n_4^2\right] = \gamma\,B$, we see that the two
statistics $\F\subA,\F\subB$ are squared sums of Gaussian unit-variance
variables, therefore they are $\chi^2$-distributed with $\dof=2$ degrees of
freedom.
The corresponding noncentrality parameters $\rho\subA^2,\rho\subB^2$, namely
\begin{equation}
  \label{eq:22}
    E\left[2\F\subA\right] = 2 + \rho\subA^2\,,\quad
    E\left[2\F\subB\right] = 2 + \rho\subB^2\,,
\end{equation}
can be expressed explicitly as
\begin{equation}
  \label{eq:23}
  \begin{aligned}
    \rho\subA^2 &= \frac{s_1^2 + s_3^2}{\gamma\,A} =
    h_0^2\gamma\left[\alpha_1\,A + \alpha_2\frac{C^2}{A} + 2\alpha_3\,C \right],\\
    \rho\subB^2 &= \frac{s_2^2 + s_4^2}{\gamma\,B} =
    h_0^2\gamma\left[\alpha_1\frac{C^2}{B} + \alpha_2\,B + 2\alpha_3\,C \right].
  \end{aligned}
\end{equation}
Note that the \emph{only} difference to the $\F$-statistic noncentrality
$\rho^2$ of Eq.~\eqref{eq:rho2} is the replacement of $B\mapsto C^2/A$ in
$\rho\subA^2$ and $A\mapsto C^2/B$ in $\rho\subB^2$.
Given that $\M$ is a positive-definite matrix, i.e., $D \equiv A\,B - C^2 > 0$
and $\alpha_1,\alpha_2>0$, this implies that
$\{\rho\subA^2,\rho\subB^2\} < \rho^2$, i.e., unsurprisingly the ``signal
power'' in the statistic is reduced by neglecting one of the two antenna
responses.
Further note that ${C^2}/{A} = B - {D}/{A}$ and ${C^2}/{B} = A - {D}/{B}$, and
therefore
\begin{equation}
  \label{eq:39}
  \begin{aligned}
    \rho^2\subA &= \rho^2 - h_0^2\gamma\,\alpha_2\frac{D}{A}\,,\\
    \rho^2\subB &= \rho^2 - h_0^2\gamma\, \alpha_1\frac{D}{B}\,.
  \end{aligned}
\end{equation}
We see that both $\rho\subA^2,\,\rho\subB^2$ approach $\rho^2$ as
$D\rightarrow 0$, so the more degenerate the antenna-pattern matrix, the less
signal power is lost.
Following Eq.~\eqref{eq:rho2Av} we can obtain the $\cosi,\psi$-averaged
noncentrality parameters as
\begin{equation}
  \label{eq:24}
  \begin{aligned}
    \av{\rho\subA^2}_{\cosi,\psi} &= \av{\rho^2}_{\cosi,\psi} - \frac{2}{5}h_0^2\gamma\,\frac{D}{A},\\
    \av{\rho\subB^2}_{\cosi,\psi} &= \av{\rho^2}_{\cosi,\psi} - \frac{2}{5}h_0^2\gamma\,\frac{D}{B}.
  \end{aligned}
\end{equation}
Given the sensitivity of $\F\subA$ and $\F\subB$ is determined by the respective
noncentrality parameters, this suggests the following practical construction for
an $\F$-statistic \emph{fallback} on degenerate segments: use either $\F\subA$
or $\F\subB$ depending on which of $\av{\rho\subA^2}$ or $\av{\rho\subB^2}$
dominates for the current sky position, which only depends on the ordering of
$A$ and $B$.
We therefore define the \emph{dominant-response} statistic $\F\subAB$:
\begin{equation}
  \label{eq:25}
  \F\subAB \equiv \begin{cases}
    \F\subA & \text{if } A > B,\\
    \F\subB & \text{otherwise},
  \end{cases}
\end{equation}
with the corresponding noncentrality parameter given by
$\rho\subAB^2 = \{\rho\subA^2 \text{ if } A > B\text{ else } \rho\subB^2\}$.

\subsubsection{Semicoherent generalization $\Fsc\subAB$}
\label{sec:semic-gener}

The coherent dominant-response statistic $\F\subAB$ defined in the previous
section can be summed semi-coherently in the same way as the standard $\F$-statistic,
i.e., following Eq.~\eqref{eq:10} we define
\begin{equation}
  \label{eq:27}
  \Fsc\subAB \equiv \sum_{\ell=1}^{N\seg} \F\subAB{}_{,\ell}\,,
\end{equation}
which is $\chi^2$-distributed with $\sc{\dof}\subAB = 2\,N\seg$ degrees of
freedom and noncentrality parameter $\sc{\rho}\subAB^2$ given by
\begin{equation}
  \label{eq:30}
    \sc{\rho}\subAB^2 = \sum_\ell \rho^2\subAB{}_{,\ell}\,.
\end{equation}
The dominant-response $\F\subAB$ statistic can be semi-coherently combined
with the standard $\F$-statistic on a per-segment basis, for example, switching to
$\F\subAB$ whenever $D$ becomes too small in a segment.
The resulting ``hybrid'' semi-coherent $\Fsc'$-statistic would be
$\chi^2$-distributed with $\sc{\dof}' = 4\,N_4 + 2\,N_2$
degrees of freedom, where $N_\dof$ is the number of segments using the
$\chi_\dof^2$-statistic, such that $N_4 + N_2 = N\seg$.
The corresponding noncentrality
parameter $\sc{\rho}^{'2}$ would be the sum of per-segment $\rho^2$
and $\rho\subAB^2$ depending on the statistic used in a given segment.

\section{Surpassing the $\Fsc$-statistic sensitivity}
\label{sec:impr-short-segm}

The new \emph{dominant-response} $\Fsc\subAB$-statistic of the previous section
was constructed as a ``fallback'' for the $\Fsc$-statistic in the degenerate
limit of an ill-conditioned antenna-pattern matrix $\M$.
As we saw in Eq.~\eqref{eq:23}, the corresponding noncentrality parameter
$\rho\subAB^2$ of $\F\subAB$ tends towards the $\F$-statistic signal
power $\rho^2$ in the limit of $D\rightarrow 0$.
However, $\Fsc\subAB$ only has \emph{two} degrees of freedom per segment
instead of the \emph{four} for the standard $\Fsc$-statistic.
Fewer degrees of freedom means a lower detection threshold at a given
false-alarm probability, which implies that there exists a range of nonzero $D$
values for which $\Fsc\subAB$ will be \emph{more sensitive} than the
$\Fsc$-statistic, even when the $\Fsc$-statistic is still perfectly well-defined
and numerically stable\footnote{This is another illustration that the
  $\F$-statistic is not optimal, as first shown in \cite{2009CQGra..26t4013P}.}.

\subsection{Estimating the maximal sensitivity gain}
\label{sec:estim-maxim-sens}

We characterize the sensitivity \cite{wette2011:_sens,2018arXiv180802459D}
of a statistic $\stat$ by the smallest required ``upper limit'' signal amplitude
$h_0^* = h_{0,p\fa}^{p\det}$ for a population of signals to reach a certain detection
probability $p\det$ at a given false-alarm probability $p\fa$.
These probabilities are defined as
\begin{align}
  p\fa &\equiv \int_{\stat\fa}^\infty \prob{\stat}{h_0=0}\,d\stat\,,  \label{eq:28a}\\
  p\det &\equiv \int_{\stat\fa}^\infty\prob{\stat}{h_0=h_0^*}\,d\stat\,,  \label{eq:28b}
\end{align}
where $\stat\fa$ denotes the detection threshold at the given false-alarm
probability $p\fa$.

One can obtain an analytic estimate for the critical signal amplitude $h_0^*$ in
the limit of a large number of segments $N\seg\gg1$, where the
$\chi^2_{\sc{\dof}}$ distributions tend toward Gaussians with the same mean and
standard deviation.
Following
\cite{krishnan04:_hough,wette2011:_sens,PrixShaltev2011..optimalStackSlide}, we
define the rescaled false-alarm threshold as
\begin{equation}
  \label{eq:32}
  \alpha \equiv \frac{\stat\fa - \sc{\dof}}{2\sqrt{\sc{\dof}}}.
\end{equation}
Using the Gaussian approximation we can solve Eq.~\eqref{eq:28a} for
\begin{equation}
  \label{eq:33}
  \alpha = \erfc^{-1}(2\,p\fa)\,,
\end{equation}
in terms of the inverse of the complementary error function $\erfc(x)$.
Similarly we define
\begin{equation}
  \label{eq:42}
  \beta \equiv -\erfc^{-1}(2\,p\det)\,,
\end{equation}
which is a monotonically increasing function of $p\det$ with $\beta>0$ for
$p\det>0.5$.
Further assuming a signal population of fixed power $\sc{\rho}^2$ (instead of
fixed amplitude $h_0$) we can express Eq.~\eqref{eq:28b} as
\begin{equation}
  \label{eq:41}
  \beta = \frac{\sc{\rho}^2 - 2\alpha\sqrt{\sc{\dof}}}{2\sqrt{\sc{\dof} + 2\sc{\rho}^2}}\,.
\end{equation}
For a given $\alpha(p\fa)$ and $\beta(p\det)$ we can solve this for the
critical signal power as
\begin{equation}
  \label{eq:34}
    \sc{\rho}^2_*(p\fa,\,p\det) = 2\alpha\sqrt{\sc{\dof}} + 4\beta^2\left(1 + \sqrt{1+Q}\right),\\
\end{equation}
with $Q \equiv (\sc{\dof} + 4\alpha\sqrt{\sc{\dof}})/(4\beta^2)$.
As shown in \cite{wette2011:_sens}, this approximation produces a biased
estimate for the critical signal amplitude $h_{0,p\fa}^{p\det}$, but we can
use it to analyze the \emph{scaling} of sensitivity with search parameters.

We can use Eq.~\eqref{eq:32} to express the respective detection thresholds in
the large-$N\seg$ limit as
\begin{equation}
  \label{eq:38}
  \frac{\Fsc\fa}{\Fsc\subAB{}\fa} = \frac{\sc{\dof} +
    2\alpha\sqrt{\sc{\dof}}}{\sc{\dof}\subAB + 2\alpha\sqrt{\sc{\dof}\subAB}}
  \overset{N\seg\gg1}{\longrightarrow} \frac{\sc{\dof}}{\sc{\dof}\subAB} = 2\,.
\end{equation}
while the critical signal power of Eq.~\eqref{eq:34} tends to
\begin{equation}
  \label{eq:36}
  \sc{\rho}_*^2 \overset{N\seg\gg1}{\longrightarrow}
  2\,(\alpha + \beta)\,\sqrt{\sc{\dof}}\,.
\end{equation}
Halving the degrees of freedom from $\sc{\dof}=4N\seg$ for the standard
$\F$-statistic to $\sc{\dof}\subAB=2N\seg$ for the dominant-response statistic
$\F\subAB$ therefore results in a maximal sensitivity gain of
\begin{equation}
  \label{eq:37}
  \frac{h_{0,\F}^*}{h_{0,\F\subAB}^*} \sim \frac{\sc{\rho}_{*,\F}}{\sc{\rho}_{*,\F\subAB}}\overset{N\seg\gg1}{\longrightarrow}
  \left(\frac{\sc{\dof}}{\sc{\dof}\subAB}\right)^{1/4} = 2^{1/4}\,,
\end{equation}
i.e., a gain of up to $\approx\SI{18.9}{\percent}$ compared to the standard
$\F$-statistic.

Note that in this derivation, we assumed $\sc{\rho}^2\subAB\approx\sc{\rho}^2$, i.e.,
$D\approx 0$ in Eq.~\eqref{eq:39}, therefore the maximal sensitivity gain would
decrease when $D$ increases, such as when using longer segments or more detectors.

\subsection{The optimal choice between $\Fsc$ and $\Fsc\subAB$}
\label{sec:estim-crit-degen}

We have seen in the previous section that when there is no loss in
noncentrality, i.e., $\sc{\rho}\subAB^2\approx\sc{\rho}^2$, the halving of the $\chi^2$
degrees of freedom results in up to $\SI{18.9}{\percent}$ sensitivity gain in
the large-$N\seg$ limit.
This argument suggests there should be a critical level of noncentrality loss
below which $\Fsc\subAB$ is more sensitive than $\Fsc$.

In order to estimate this transition we use Eq.~\eqref{eq:36} to express
$\beta = \sc{\rho}^2/\sqrt{\sc{\dof}} - 2\alpha$, where $\beta$ is a monotonic
function of detection probability $p\det$.
Therefore the more sensitive statistic is characterized by a higher
$\sc{\rho}^2/\sqrt{\sc{\dof}}$.
This can be reformulated into the condition
\begin{equation}
  \label{eq:46}
  \sc{\mu}\subAB \equiv \frac{\sc{\rho}^2 - \sc{\rho}\subAB^2}{\sc{\rho}^2} <
  \frac{\sqrt{2} - 1}{\sqrt{2}} \approx0.29\,,
\end{equation}
in terms of the \emph{noncentrality mismatch} $\sc{\mu}\subAB$.
Therefore up to a loss of $\sc{\mu}\subAB\lesssim \SI{29}{\percent}$ in
noncentrality there is still a net sensitivity gain for the
dominant-response statistic $\Fsc\subAB$ due to the reduction in degrees of
freedom.

This condition is not practically usable as it depends on the unknown signal
amplitude parameters $\cosi$ and $\psi$, but we can use the averages of
Eq.~\eqref{eq:24} to obtain an estimate
\begin{equation}
  \label{eq:47}
  \sc{m}\subAB\equiv\av{\mu\subAB}_{\cosi,\psi} \sim \frac{\sum_\ell
    \frac{D_\ell}{\max(A,B)_\ell}}{\sum_\ell (A_\ell + B_\ell)}\,,
\end{equation}
where we approximated the average of the fraction as the fraction of averages,
which is generally biased but numerically turns out to be a relatively good
approximation in this case.
Therefore we expect the dominant-response $\F\subAB$-statistic to be more
sensitive than the $\F$-statistic when
\begin{equation}
  \label{eq:55}
  \sc{m}\subAB\lesssim 0.29\,.
\end{equation}
For sufficiently-short segments, this criterion will be satisfied essentially
over the whole sky, favoring $\Fsc\subAB$, while for long-enough
segments it will not be satisfied anywhere, favoring $\Fsc$, with an
intermediate regime of segment lengths where the optimal choice will depend on
the sky position.

\section{Numerical results}
\label{sec:numerical-tests}

In this section, we numerically test the theoretical predictions derived in the
previous sections and further characterize the performance of the new statistic
$\Fsc\subAB$.
We use two complementary numerical methods for sensitivity estimation in order
to cross-check and validate the implementations and results, namely
\begin{enumerate}
\item Direct numerical integration of the known $\chi^2$ distributions for the
  two statistics $\Fsc$ and $\Fsc\subAB$, using the explicit noncentrality
  expressions for $\sc{\rho}^2$ and $\sc{\rho}^2\subAB$ given in
  Eq.~\eqref{eq:rho2} and Eq.~\eqref{eq:23}, respectively.
  This is essentially the sensitivity-estimation method described in in
  \cite{wette2011:_sens,2018arXiv180802459D}.

\item Sampling \emph{synthesized} statistic values (per segment) by drawing noise-realizations
  for the four Gaussian variates $n_\mu$ with covariance matrix $\M_{\mu\nu}$ of
  Eq.~\eqref{eq:20}, and adding signal contributions
  $s_\mu = \M_{\mu\nu}\,\A^\nu$ for $\A^\mu$ drawn from their priors.
  Both the standard $\F$-statistic as well as $\F\subAB$ are fully determined by
  the four numbers $x_\mu = n_\mu + s_\mu$, as seen in Eq.~\eqref{eq:15} and
  Eq.~\eqref{eq:11}, \eqref{eq:17}, respectively.
  This synthetic sampling method has been used previously in
  \cite{2009CQGra..26t4013P,keitel_f-statistic_2012}.
\end{enumerate}
Both these methods are substantially faster than full injection-and-recovery
simulations and ``cleaner'' in the sense of not being affected by
noise-estimation biases and intrinsic implementation losses when computing
$\Fsc$-statistics on actual data.
Therefore, these methods allow us to explore larger parameter spaces and
characterize the intrinsic properties of these statistics.
Compared to direct integration, sampling of synthetic statistic values is more
limited due to the finite number of samples generated in a given amount of time,
resulting in lower accuracy and limitations on how small false-alarm
probabilities $p\fa$ can be simulated.

For these tests we assume perfect Gaussian noise and data without gaps.  For the
sake of example, sensitivity is characterized at an ``upper limit'' confidence
level of $p\det=\SI{90}{\percent}$ and for a \emph{single-template} false-alarm
level of $p\fa=\num{e-10}$.
This choice would correspond to an overall false-alarm level of
$\sim\SI{1}{\percent}$ over a parameter-space region containing $\sim\num{e8}$
(independent) templates, which would be a reasonably realistic choice for typical
wide-parameter-space searches.

The sensitivity of a search can be conveniently characterized by its
\emph{sensitivity depth} \cite{2014arXiv1410.5997B,2018arXiv180802459D}, defined
as
\begin{equation}
  \label{eq:53}
  \depth_{p\fa}^{p\det}\equiv \frac{\sqrt{\S}}{h_{0,p\fa}^{p\det}}\,,
\end{equation}
in terms of the overall noise power spectral density $\S$ of Eq.~\eqref{eq:54}
and the sensitivity amplitude $h_{0,p\fa}^{p\det}$ introduced in
Sec.~\ref{sec:estim-maxim-sens}.
This definition scales out the noise-floor dependence and better characterizes
the intrinsic method sensitivity.

\subsection{Example: segments of $T\seg=\SI{900}{\second}$ over 6 months}
\label{sec:example-short-segm}

We consider the following example setup\footnote{Based on a recent all-sky
  binary search \cite{covas_constraints_2022} on O3a data, which used
  $N\seg=\num{16966}$ segments (with gaps) of $T\seg=\SI{900}{\second}$.} to
illustrate the performance of the dominant-response statistic $\Fsc\subAB$
compared to the standard $\Fsc$-statistic: an all-sky search on 6 months of data
without gaps using $N\seg=\num{17280}$ segments of length
$T\seg=\SI{900}{\second}$.

The detection thresholds for the two statistics are found as
$\Fsc\fa \approx \num{6.4e+04}$ and $\Fsc\subAB{}\fa \approx \num{3.3e+04}$, with a ratio of
$\approx1.97$, very close to the theoretical prediction of
Eq.~\eqref{eq:38}.

We can numerically estimate the sensitivity depths for the two statistics as
shown in Table~\ref{tab:example-depths}, for one (H1), two (H1+L1) or three
detectors (H1+L1+V1).
\begin{table}[tbp]
  \centering
  \begin{tabular}{c|ccr}
detectors & $\depth(\Fsc)$ & $\depth(\Fsc\subAB)$ & gain\\\hline
H1 & $\num{18.3}$ & $\num{21.7}$ & $\SI{18.6}{\percent}$ \\
H1,L1 & $\num{26.3}$ & $\num{30.5}$ & $\SI{16.2}{\percent}$ \\
H1,L1,V1 & $\num{32.1}$ & $\num{35.0}$ & $\SI{9.0}{\percent}$ \\
\end{tabular}

  \caption{Sensitivity depth $\depth$ (in units of \udepth) for the standard
    $\Fsc$-statistic and the dominant-response $\Fsc\subAB$-statistic
    at $p\fa=\num{e-10}$ and $p\det=\SI{90}{\percent}$.
    The search setup consists of $N\seg=\num{17280}$ segments of
    $T\seg=\SI{900}{\second}$, for different sets of detectors.
  }
  \label{tab:example-depths}
\end{table}
We see that for a single detector the sensitivity gain of $\Fsc\subAB$ versus $\Fsc$
is close to the theoretical maximum of $\SI{18.9}{\percent}$, while
increasing the number of detectors reduces the gain, down to about
$\SI{9}{\percent}$ for three detectors.
This behavior stems from the short-segment antenna pattern being more degenerate
for a single detector (i.e., $D$ is closer to zero), keeping the noncentrality
loss of $\sc{\rho}\subAB^2$ versus $\sc{\rho}^2$ small while using more
detectors reduces the degeneracy and thereby increases the mismatch
$\sc{m}\subAB$.

\begin{figure}[htbp]
  \centering
  \includegraphics[width=\columnwidth]{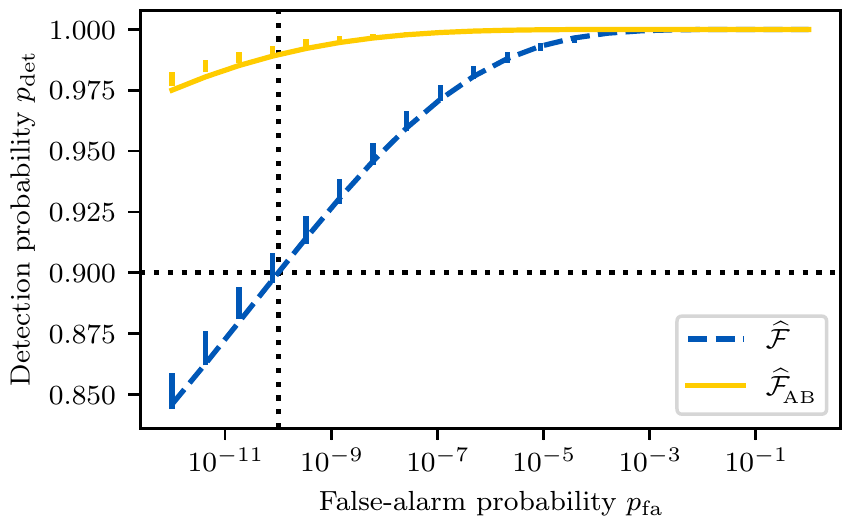}
  \caption{Receiver-operator characteristic for the dominant-response
    $\Fsc\subAB$-statistic and the standard $\Fsc$-statistic, on a signal
    population of fixed depth $\depth_{\num{e-10}}^{\SI{90}{\percent}}(\Fsc)$,
    using two different numerical methods: direct integration (solid lines)
    and synthetic statistic sampling (error bars showing $90\%$-confidence
    regions).
    The search setup consists of $N\seg=\num{17280}$ segments of
    $T\seg=\SI{900}{\second}$ for two detectors (H1+L1).
  }
  \label{fig:rocs}
\end{figure}

\begin{figure}[htbp]
  \includegraphics[width=\columnwidth]{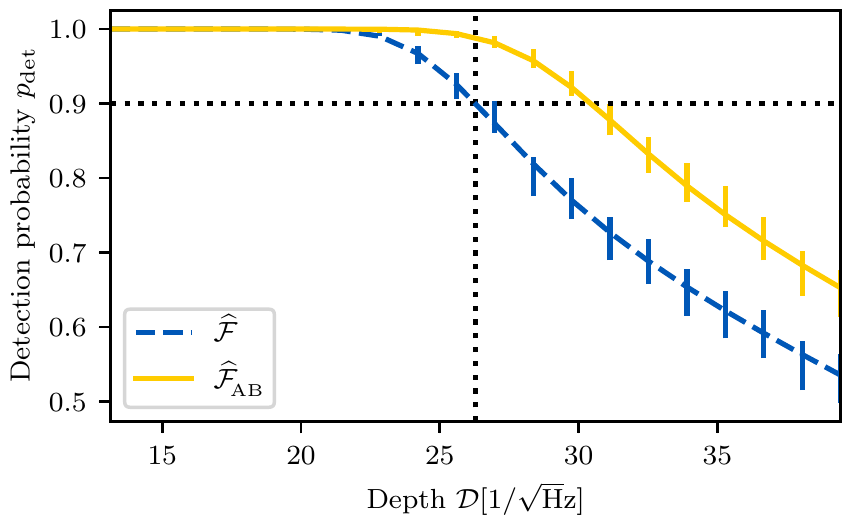}
  \caption{Efficiency curves for the dominant-response $\Fsc\subAB$-statistic and
    the standard $\Fsc$-statistic, at a (single-template) false-alarm level of
    $p\fa=\num{e-10}$, using two different numerical methods: direct integration
    (solid lines) and synthetic statistic sampling (error bars showing
    $90\%$-confidence regions).
    The search setup consists of $N\seg=\num{17280}$ segments of
    $T\seg=\SI{900}{\second}$ for two detectors (H1+L1).  }
  \label{fig:efficiency}
\end{figure}

Figure~\ref{fig:rocs} shows the receiver-operator characteristic, i.e., $p\det$
versus $p\fa$, for two detectors (H1+L1) and signals at fixed depth
$\depth_{\num{e-10}}^{\SI{90}{\percent}}(\Fsc)$ of Table~\ref{tab:example-depths}.
At the false-alarm level of $p\fa=\num{e-10}$, the detection probability of the
dominant-response statistic $\Fsc\subAB$ is about $\SI{8}{\percent}$ higher than
that of the $\Fsc$-statistic.
Figure~\ref{fig:efficiency} shows the \emph{efficiency curve}, i.e., $p\det$
versus signal strength quantified in terms of the depth $\depth$.

\subsection{Sensitivity gain versus segment length $T\seg$}
\label{sec:sens-gain-vers}

\begin{figure}[htbp]
  \centering
  \includegraphics[width=\columnwidth]{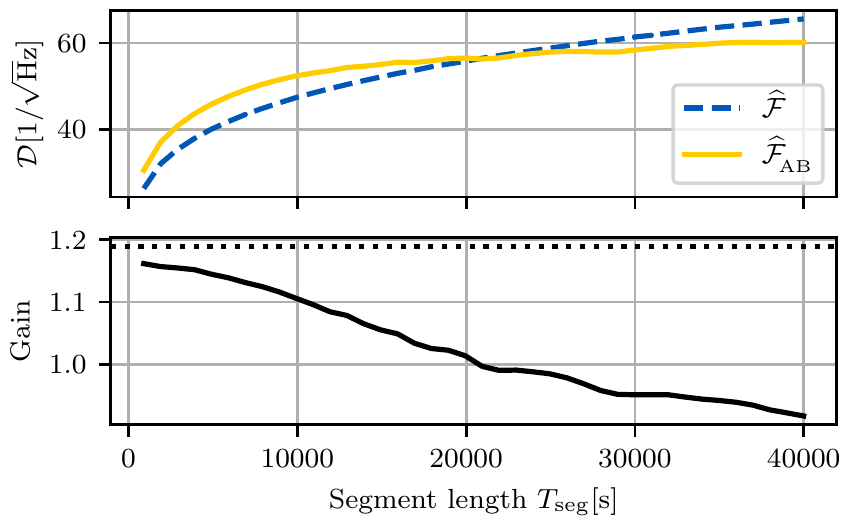}
  \caption{Sensitivity depth $\depth_{\num{e-10}}^{\SI{90}{\percent}}$ and
    relative sensitivity gain as a function of segment length $T\seg$ for the
    dominant-response $\Fsc\subAB$-statistic versus the standard $\Fsc$-statistic.
    The dotted horizontal line indicates the theoretical maximal sensitivity
    gain of $\SI{18.9}{\percent}$.
    The search setup spans \SI{6}{months} using two detectors (H1+L1).
  }
  \label{fig:gain-vs-tseg}
\end{figure}

An interesting question of practical importance is up to which 'critical'
segment length $T\seg^*$ the dominant-response $\Fsc\subAB$-statistic performs
better than the $\Fsc$-statistic.
The answer to this question is shown in Fig.~\ref{fig:gain-vs-tseg}, for a
single-detector (H1) and two-detector (H1+L1) all-sky search.
We see that for both H1 and H1+L1 searches, the transition point is at about
$T\seg^*\sim \SI{20000}{\second}$, while adding Virgo (H1+L1+V1) would bring
this down to about $T\seg^*\sim\SI{15000}{\second}$ (not shown here).

The underlying reason for the decreasing gains of $\Fsc\subAB$ lies in the
increasing noncentrality mismatch $\sc{m}\subAB$, which will also depend on the
sky position.
Therefore, searches directed at a single sky position will have different
transition segment lengths dependent on that sky position.

\begin{figure}[htbp]
  \centering
  \includegraphics[width=\columnwidth]{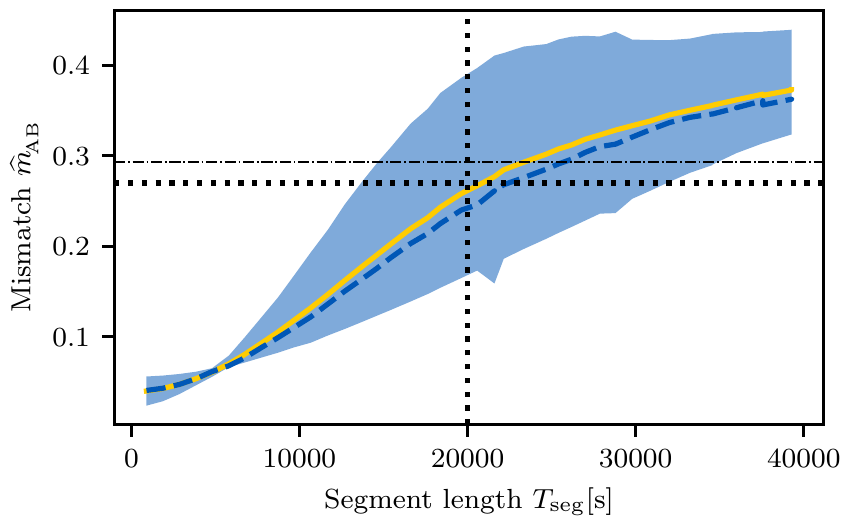}
  \caption{Distribution (over the sky) of the noncentrality mismatch $\sc{m}\subAB$ as a
    function of segment length $T\seg$ for a two-detector search setup (H1+L1).
    The solid line indicates the mean mismatch, the dashed
    line shows the median, and the band denotes the range from the 10th to the
    90th percentile.
    The horizontal dot-dashed line corresponds to the theoretical
    estimate $\sc{m}\subAB^*\approx0.29$ of Eq.~\eqref{eq:55} for the critical
    mismatch, while the dotted vertical line indicates the observed transition
    segment length $T\seg^*\sim\SI{20000}{\second}$ of
    Fig.~\ref{fig:gain-vs-tseg}.
    The resulting \emph{empirical} estimate for the critical mismatch is
    $\sc{m}\subAB^*\sim 0.27$ (dotted horizontal line).
  }
  \label{fig:mis-vs-tseg}
\end{figure}

To partially address this question, in Fig.~\ref{fig:mis-vs-tseg} we plot the
mismatch distribution (for H1+L1) versus segment length.
This shows that, somewhat consistently with Fig.~\ref{fig:gain-vs-tseg} and the
theoretical estimate Eq.~\eqref{eq:55}, the mean noncentrality mismatch crosses
the theoretical estimate of the 'critical' mismatch of $\sc{m}\subAB^*\sim0.29$
(dashed horizontal line) at about $T\seg\sim\SI{22000}{\second}$, while at the observed critical
segment length of $T\seg\sim\SI{20000}{\second}$ in Fig.~\ref{fig:mis-vs-tseg}
the corresponding \emph{empirical} critical mismatch would be about
\begin{equation}
  \label{eq:56}
  \sc{m}\subAB^*\sim 0.27\,.
\end{equation}
From this plot, we would also predict $\Fsc\subAB$ to be more sensitive in every
sky position below $T\seg\lesssim \SI{15000}{\second}$, while $\Fsc$ should be
more sensitive in every sky position above $T\seg\gtrsim\SI{35000}{\second}$.
For intermediate segment lengths, the choice of the optimal statistic will be a
function of the sky position.

\begin{figure*}[htbp]
  \centering
  \includegraphics[width=\columnwidth]{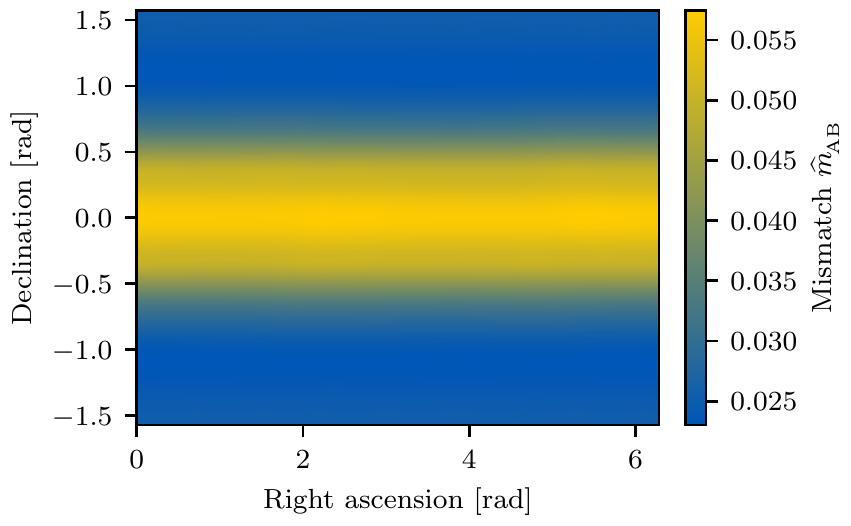}
  \includegraphics[width=\columnwidth]{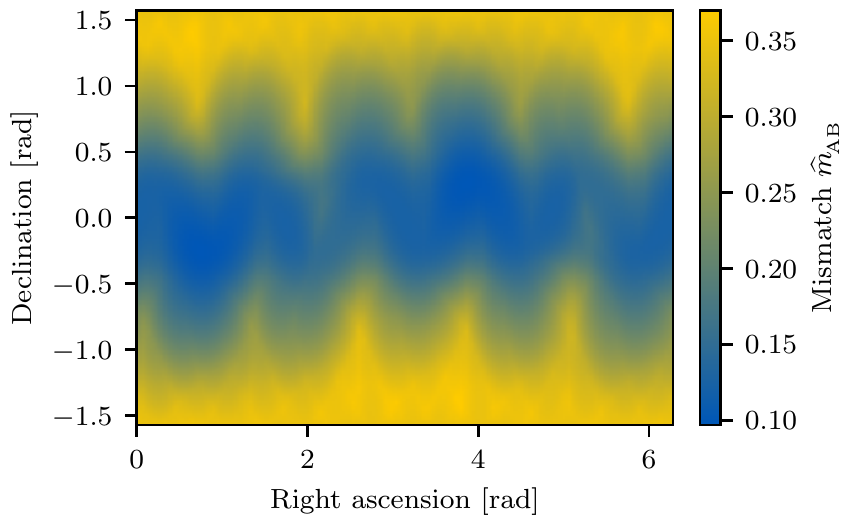}
  \caption{Noncentrality mismatch $\sc{m}\subAB$ as a function of sky position
    for a two-detector search setup (H1+L1) spanning one day,
    using $N\seg=96$ segments of $T\seg=\SI{900}{\second}$ duration
    (\emph{left plot}) and $N\seg=6$ segments of $T\seg=\SI{17100}{\second}$
    duration \emph{(right plot)}.
  }
  \label{fig:mis-over-sky}
\end{figure*}

\subsection{Sensitivity gain versus mismatch $\sc{m}\subAB$}
\label{sec:sens-gain-vers-1}

From earlier theoretical considerations in Sec.~\ref{sec:estim-crit-degen} as
well as the empirical results in the previous section, we expect the
noncentrality mismatch $\sc{m}\subAB$ to be the intrinsic factor determining the
gain of $\Fsc\subAB$ versus $\Fsc$.
This mismatch is a function not only of the segment setup and detectors but also
of the sky position, as illustrated in Fig.~\ref{fig:mis-over-sky}.  Here we see
the noncentrality mismatch for a two-detector (H1+L1) setup over one day (the
antenna patterns are periodic over a day) for two different segment lengths,
$T\seg=\SI{900}{\second}$ and $T\seg=\SI{17100}{\second}$.
This example illustrates that the mismatch can be highest at the poles or the
equator, depending on the detailed setup (detectors, segment lengths), and
increases for longer segments.

Following the discussion in Sec.~\ref{sec:estim-crit-degen} and especially
Eq.~\eqref{eq:55}, we expect the critical mismatch value to be around
$\sc{m}\subAB^*\approx 0.29$, while empirical results in
Fig.~\ref{fig:mis-vs-tseg} and Eq.~\eqref{eq:56} suggest a slightly lower
critical value of about $\sc{m}\subAB\sim0.27$.

\begin{figure}[htbp]
  \centering
  \includegraphics[width=\columnwidth]{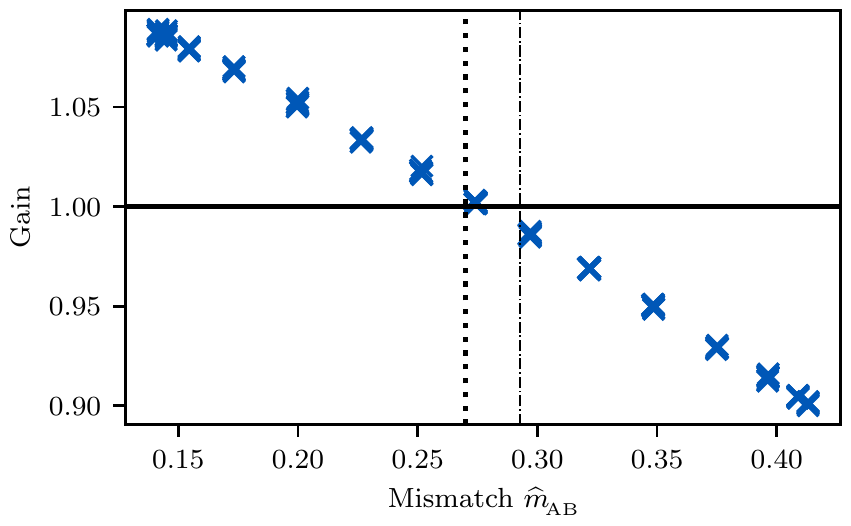}
  \caption{Sensitivity gain of the dominant-response $\Fsc\subAB$-statistic over
    the standard $\Fsc$-statistic for single-sky-position searches as a function
    of the corresponding noncentrality mismatch $\sc{m}\subAB$ in the sky position.
    The vertical dot-dashed line corresponds to the theoretical estimate
    $\sc{m}\subAB^*\approx0.29$ of Eq.~\eqref{eq:55}, while the vertical dotted
    line indicates the empirical critical mismatch of
    $\sc{m}\subAB^*\sim 0.27$ estimated in Fig.~\ref{fig:mis-vs-tseg}.
    The search setup consists of $N\seg=\num{910}$ segments of
    $T\seg=\SI{17100}{\second}$ duration using two detectors (H1+L1).
  }
  \label{fig:gain-vs-mis}
\end{figure}

We can test these predictions by plotting the sensitivity gain as a function of
noncentrality mismatch $\sc{m}\subAB$ (by varying the sky position), where we
chose an example near the transition, namely $T\seg=\SI{17100}{\second}$.
The result is shown in Fig.~\ref{fig:gain-vs-mis}, for a two-detector setup
(H1+L1) and a \SI{6}{months} total observation span.
These results confirm that indeed the sensitivity gain correlates strongly with
the noncentrality mismatch and that the transition does happen close to the
theoretically predicted value, albeit with the empirical prediction of
$\sc{m}\subAB^*\sim0.27$ being closer to the truth.

\section{Conclusions}
\label{sec:conclusions}

In this paper, we have shown that the standard $\F$-statistic becomes singular
for very short segments due to the degeneracy of the antenna-pattern matrix.
This observation lead us to construct a well-behaved ``fallback'' statistic,
referred to as the \emph{dominant-response} $\F\subAB$-statistic.
Somewhat surprisingly, however, this new statistic turns out to be \emph{more
  sensitive} than the $\F$-statistic by up to $\approx\SI{19}{\percent}$, even
outside the degeneracy region of the $\F$-statistic, when semi-coherently
combining of a large number of short segments.
We have characterized the new $\F\subAB$-statistic analytically and numerically
and shown that it is more sensitive than the $\F$-statistic for segments shorter
than $\Tspan\lesssim$ \SIrange{15000}{20000}{\second} (depending on the number
of detectors used).

We have further shown that the sensitivity gain is determined by the
noncentrality mismatch $\sc{m}\subAB$ of Eq.~\eqref{eq:47}, and theoretical and
empirical estimates place the transition at about $\sc{m}\subAB\sim0.27$, below
which $\F\subAB$ is more sensitive than $\F$.

The new detection uses the same signal phase model as the $\F$-statistic and is
therefore expected \cite{prix06:_searc} to require essentially the same number
of templates and computing cost for a given parameter space.

Future work will focus on re-analyzing this limit in a Bayesian framework
following the approach of \cite{2009CQGra..26t4013P}.

\begin{acknowledgments}
  This work has utilized the ATLAS computing cluster at the MPI for
  Gravitational Physics Hannover.
\end{acknowledgments}

\appendix

\section{Constant-response statistic $\F\subphi$}
\label{subsubsec:constantantenna}

As an alternative construction to the dominant-response statistic $\F\subAB$
introduced in Sec.~\ref{sec:domin-polar-f}, it is interesting to consider
another special case obtained by assuming a constant amplitude response, i.e.,
completely neglecting the antenna-pattern modulation, and define a template
\begin{equation}
  \label{eq:29}
  \begin{aligned}
    h^X\subphi(t) &= \A\,\cos(\phi^X(t) + \phi_0)\\
    &= \A\subc\,\cos\phi^X(t) + \A\subs\,\sin\phi^X(t)\,,
  \end{aligned}
\end{equation}
where $\A\subc \equiv \A\,\sin\phi_0$ and $\A\subs \equiv \A\,\cos\phi_0$.
Using this template family the log-likelihood of Eq.~\eqref{eq:loglikelihood}
takes the form
\begin{equation}
  \label{eq:31}
  \ln\mathcal{L} = \A\subs\,x\subs + \A\subc\,x\subc - \frac{\gamma}{2}\,\A^2\,,
\end{equation}
where we defined
\begin{equation}
  \label{eq:40}
  x\subc \equiv (x|\cos\phi)\,,\quad
  x\subs \equiv (x|\sin\phi)\,,
\end{equation}
in terms of the (multi-detector) scalar product $(x|y)$ of Eq.~\eqref{eq:4}.
Maximizing this log-likelihood over the two unknown amplitudes
$\A\subc,\,\A\subs$ yields $\A\subsc\mle=\gamma^{-1}\,x\subsc$, and substituting
back results in the partially-maximized likelihood
\begin{equation}
  \label{eq:44}
  2\F\subphi \equiv 2\ln\mathcal{L}\mle = \gamma^{-1}\left(x\subc^2 + x\subs^2\right),
\end{equation}
defining the \emph{constant-response} $\F\subphi$-statistic.
Note that $x\subc^2 + x\subs^2$ in the single-detector case is precisely the
Fourier power of $x(t)$ demodulated into the source frame.

In the short-segment limit $T\seg\rightarrow0$ for a single detector this would
tend exactly to the dominant-response $\F\subAB$-statistic, as in this limit
$a(t)\rightarrow a_0$, $b(t)\rightarrow b_0$, and therefore
$A \rightarrow a_0^2$, $B\rightarrow b_0^2$ and also
$x_\mu\rightarrow \{a_0,b_0\}\,x_{\mathrm{s,c}}$.

Similarly to Sec.~\ref{sec:domin-polar-f} one can show that this statistic
follows a $\chi^2$ distribution with $\dof=2$ degrees of freedom, namely
\begin{equation}
  \label{eq:35}
  E[2\F\subphi] = 2 + \rho\subphi^2\,,
\end{equation}
with a noncentrality parameter
\begin{equation}
  \label{eq:45}
  \rho\subphi^2 \equiv \gamma^{-1}(s\subc^2 + s\subs^2)\,,
\end{equation}
in terms of the signal projections $s\subsc \equiv (s|\{\sin,\cos\}\phi)$.
Using the scalar product of Eq.\eqref{eq:4} with the basis functions
of Eq.~\eqref{eq:basisfunc}, we find
\begin{equation}
  \label{eq:50}
  \begin{aligned}
    s\subc &= \gamma\,\left(\A^1\,\avS{a} + \A^2\,\avS{b}\right)\,,\\
    s\subs &= \gamma\,\left(\A^3\,\avS{a} + \A^4\,\avS{b}\right)\,,
  \end{aligned}
\end{equation}
using the noise-weighted multi-detector average defined in Eq.~\eqref{eq:1}.
Combining this with Eq.~\eqref{eq:51} we can express the constant-response
noncentrality as
\begin{equation}
  \label{eq:52}
  \rho\subphi^2 = h_0^2\gamma\,\left[\alpha_1\,\avS{a}^2 + \alpha_2\avS{b}^2 + 2\alpha_3\avS{a}\avS{b}\right]\,,
\end{equation}
which has an uncanny formal resemblance to the $\F$-statistic signal power
expression of Eq.~\eqref{eq:rho2}, except that the mean squares of
antenna-pattern functions of Eq.~\eqref{eq:APC} are now replaced by squared
means, i.e., $\avS{a^2}\mapsto\avS{a}^2$, $\avS{b^2}\mapsto\avS{b}^2$ and
$\avS{a b}\mapsto\avS{a}\avS{b}$.
This observation implies that for a single detector in the short-segment limit,
this statistic will again have the same noncentrality as the $\F$-statistic with
only half the degrees of freedom, gaining the same $\approx\SI{18.9}{\percent}$ in
sensitivity as the dominant-response $\F\subAB$-statistic of
Sec.~\eqref{sec:domin-polar-f}.

For longer observation times as well as multiple detectors, however, the
antenna-pattern function can cancel out in the averages, resulting in
potentially significant sensitivity losses.  Especially for non-aligned
detectors, the multi-detector averages can cancel strongly, potentially making
$\F\subphi$ more suitable as a ``veto'' rather than a detection statistic.

\bibliography{bib}

\end{document}